\documentclass{elsarticle}

\usepackage{url,lineno,microtype,subcaption}
\usepackage{amssymb}
\usepackage{gensymb}
\usepackage[version=4]{mhchem}
\usepackage{graphicx}
\usepackage{amsmath, mathtools}
\usepackage{natbib}
\usepackage[colorlinks=true,linkcolor=black, citecolor=blue, urlcolor=blue]{hyperref}
\usepackage{appendix}

\newcommand{\puo}{PuO$_{2}$}

\begin{document}

\title{Investigation of process history and phenomenology of plutonium oxides using vector quantizing variational autoencoder} 

\author[add1]{Connor Hainje}
\author[add1]{Cody Nizinski}
\author[add1]{Shane Jackson}
\author[add1]{Richard Clark}
\author[add1]{Forrest Heller}
\author[add1]{Ian Schwerdt}
\author[add1]{Edgar Buck}
\author[add1]{David Meier}
\author[add1]{Alexander Hagen}
\ead{alexander.hagen@pnnl.gov}

\address[add1]{Pacific Northwest National Laboratory, Richland, WA 99352, USA}

\begin{abstract}

Accurate, high throughput, and unbiased analysis of plutonium oxide particles is needed for analysis of the phenomenology associated with process parameters in their synthesis.  Compared to qualitative and taxonomic descriptors, quantitative descriptors of particle morphology through scanning electron microscopy (SEM) have shown success in analyzing process parameters of uranium oxides.  Among other candidates, a neural network called a Vector Quantizing Variational Autoencoder (VQ-VAE) has shown the ability to quantitatively describe particle morphology to attain $>$85\% accuracy in identifying uranium oxide processing routes. We utilize a VQ-VAE to quantitatively describe plutonium dioxide (\puo{}) particles created in a designed experiment and investigate their phenomenology and prediction of their process parameters. \puo{} was calcined from Pu (III) oxalates that were precipitated under varying synthetic conditions that related to concentrations, temperature, addition and digestion times, precipitant feed, and strike order; the surface morphology of the resulting \puo{} powders were analyzed by SEM. A pipeline was developed to extract and quantify useful image representations for individual particles with the VQ-VAE, then further reduce the dimensionality of the feature space using a bottlenecking neural network fit to perform multiple classification tasks simultaneously. The reduced feature space could predict process parameters with greater than 80\% accuracies for some parameters with a single particle. They also showed utility for grouping particles with similar surface morphology characteristics together. Both the clustering and classification results reveal valuable information regarding which chemical process parameters chiefly influence the \puo{} particle morphologies: strike order and oxalic acid feedstock. Doing the same analysis with multiple particles was shown to improve the classification accuracy on each process parameter over the use of a single particle, with statistically significant results generally seen with as few as four particles in a sample.

\end{abstract}


\date{\today}

\maketitle

\section{Introduction}

Plutonium oxalate precipitates have historically been an important intermediate material for the production of plutonium oxides, PuO$_{2}$. Plutonium oxalate can be precipitated from Pu (III) nitrate solutions (Eq.~\ref{III}) or Pu (IV) nitrate solutions (Eq.~\ref{IV}) \cite{peterson2019engineering}. High-temperature calcination of either Pu oxalate converts the precipitates to plutonium oxide, \puo{}. Not only have the \puo{} materials produced by different precipitation routes observed to have unique properties \cite{moseley1965properties}, but the precipitation conditions within a single route can also greatly affect the material properties \cite{facer1954precipitation, smith1976effect, burney1984controlled, hagan1980plutonium}.

\begin{equation}
    \ce{2Pu(NO3)3 + 3H2C2O4 + 10H2O -> Pu2(C2O4)3.10H2O + 6HNO3}
    \label{III}
\end{equation}
\begin{equation}
    \ce{Pu(NO3)4 + 2H2C2O4 + 6H2O -> Pu2(C2O4)2.6H2O + 4HNO3}
    \label{IV}
\end{equation}

Well-designed experiments can help understand the complex relationships between the synthesis parameters and the measurable properties of the resulting material without requiring samples to be synthesized exhaustively under all possible combinations of parameters \cite{anderson2015design}, which is especially important when considering the unique hazards and high costs of working with and disposing of transuranic elements. Earlier statistical design of experiments (DOE) for plutonium oxides sought to control a singular material property such as the the particle size of powders for fuel production \cite{smith1976effect, burney1984controlled} or the filterability of precipitates \cite{hagan1980plutonium}. The particle size studies first investigated several plutonium precipitation routes \cite{smith1976effect} before studying the Pu (III) oxalate route in greater depth \cite{burney1984controlled}.

Particle size and, more broadly, the morphological characteristics of nuclear materials hold valuable insights relating to the processes that created them \cite{mayer2013nuclear}. Process factors including but not limited to precipitation temperature, solution concentrations, strike order, and digestion times can each affect the morphology of a material \cite{mayer2013nuclear, anderson2015design}. The morphological characteristics of lanthanide and actinide materials have been an active topic of research in recent years, with studies focusing on modeling crystal growth \cite{garrett2018first, tamain2013crystal}, investigating structural properties \cite{runde2009directed, abraham2014actinide}, and relating chemical synthesis parameters to material properties \cite{olsen2017quantifying, tyrpekl2017alterations, tyrpekl2019cerium, thompson2021nuclear}.

To describe and characterize the surface morphologies of nuclear materials, a variety of computational image analysis have been explored \cite{nizinski2021computational}. Tamasi et al. \cite{tamasi2016lexicon} proposed a multi-step lexicon to provide a consistent vocabulary when describing the shape and texture characteristics of individual particles and aggregates; while the lexicon has seen adoption with uranium oxide powders, the data produced by this method remains fairly inconsistent and subjective. Quantitative particle analysis for nuclear materials has typically leveraged Los Alamos National Laboratory's MAMA software \cite{ruggiero2014mama} to extract size and shape features; these features have been paired with various single- and multi-variate analysis techniques to measure how changes to a material's process history affect the surface morphology \cite{olsen2017quantifying, lesiak2021characterization, burr2021overview}. More recently, convolutional neural networks (CNN) have been explored for classifying the process histories of uranium oxides from scanning electron microscopy (SEM) images. The Multi-Input Single-Output (MISO) \cite{ly2020determining} method utilized supervised learning and several ResNet50 \cite{he2016deep} models to achieve accuracies $>$95\% for twelve-way classification of uranium oxides synthesized by different processing routes and imaged by SEM at four magnifications.

An alternative CNN methodology was created by Girard and Hagen, using a vector-quantized variational autoencoder (VQ-VAE) to generate quantitative descriptions of particle images, and using those descriptions to classify different uranium synthetic pathways with unique surface morphologies \cite{girard2021uranium}. The VQ-VAE was initially developed for image generation \cite{van2017neural}.  In lieu of a continuous embedding space, the VQ-VAE simultaneously learns a set number of codebook embedding vectors and the encoder weights to minimize the distance between the outputs of the encoder and the embedding vectors.  The encoder-decoder are trained using reconstruction loss for the images in question.  Images can then be represented by the lower dimensional space of histograms of their codebook vectors.  Hierarchical VQ-VAE improves the resolution of the reconstruction by adding additional latent maps \cite{razavi2019generating} with a smaller, so-called top embedding capturing global phenomena of the image while the usual bottom embedding captures the local structures.  The reduced-dimension representations of a two-encoder hierarchical VQ-VAE were shown to be very useful for classification, achieving $>$85\% accuracy on twelve-way classification with smaller data size requirements \cite{girard2021uranium} than that of other CNNs.

Recent work has also been concerned with this inverse modeling problem for designed experiments, i.e., estimating the synthesis parameters from the observed chemical, physical, and morphological properties of \puo{} \cite{anderson2015design, lewis2018comparing, ausdemore2022probabilistic}. These efforts have sought to further develop techniques and methodologies capable of obtaining robust, multi-faceted morphological information relating to nuclear material synthetic process. Each of these studies \cite{anderson2015design, lewis2018comparing, ausdemore2022probabilistic} used particle size and shape data. Lewis et al. \cite{lewis2018comparing} applied a diverse collection of statistical methodologies for both the forward modeling and inverse problems, noting the benefit of comparing multiple statistical methods to diagnose and evaluate the predictive capabilities of the models on a dataset. Anderson-Cook et al. \cite{anderson2015design} noted the challenges of using surface morphology data -- specifically size and shape features extracted from micrographs -- for design of experiment responses, concluding that more robust representations for surface morphologies are needed.

\section{Materials and methods}

\subsection{Synthesis of PuO$_2$}

\puo{} was synthesized using Pacific Northwest National Laboratory's bench-scale plutonium oxide processing capability \cite{lumetta2019puo2}. To assess how processing parameters of the Pu (III) oxalate precipitation route affect the physical, chemical, and morphological characteristics of the resulting \puo{} a parametric design of experiment study was developed. Seven variables were evaluated as part of the statistical DOE: Pu concentration, nitric acid concentration, strike direction, oxalic acid feed, addition time, digestion time, and precipitation temperature. Values for these parameters are compiled in Table~\ref{table:enc}. In total, seventy-six synthetic runs were performed, including several with duplicated experimental conditions so that the consistency of \puo{} processing and analytical methods could be evaluated.

The precipitation vessel design, agitator design and speed, calcination conditions, and storage conditions were kept constant for all runs. After washing the Pu (III) oxalate precipitates, they were recovered by vacuum filtration and dried on the filter before being recovered. The Pu (III) oxalates were calcined to \puo{} in a muffle furnace by heating to 650\degree C with a dwell time of two hours and fixed temperature ramping profiles.


\subsubsection{SEM data collection}

To obtain representative samples, coning and quartering \cite{dunn1997exercise} was used to partition the bulk powders for SEM and other analytical techniques. To mount samples for SEM analysis, 10 mg of \puo{} powder was dispersed in 3 mL of isopropyl alcohol (IPA) and mixed for 1 minute on a vortex mixer at 2,000 rpm. A micropipette was used to immediately transfer a 5 $\mu$L droplet of the suspension to the center of the surface of a 25.4 mm aluminum planchette covered with an Al-backed carbon tape; once the IPA had evaporated, the stub was inverted and vigorously tapped to remove any loose particulates. The sample stub exterior and the aluminum pan were confirmed to be free of removable contamination, tap tested in a fume hood, and then transferred to the SEM for imaging. Sputter coating was not performed as a part of the sample preparation.

The developed mounting method satisfied several essential criteria. Particles were sufficiently dispersed by the IPA suspension such that the mounted particulates had minimal clumping and overlap on the SEM stub. There were a sufficient quantity of particles that allowed for quantitative analysis by segmentation and for the machine learning methods described in this work. The mounted particles are believed to be representative of the bulk powder. Finally, SEM samples were prepared in way that was free of contamination and stayed under the benchtop limits for work with plutonium.

An FEI Quanta 250 field emission gun (FEG)-SEM was used to analyze the \puo{} samples using both the backscatter electron (BSE) and secondary electron (SE) imaging modes. Magnifications of 1,000$\times$, 1,500$\times$, and 2,500$\times$ were used. To avoid imaging the same particle multiple times, a rastering pattern progressing from the top left of the stub towards the bottom right that left sufficient space between each collected micrograph was implemented. The SEM analyses were consistently performed using a 15.0 kV beam voltage, a spot size of 5.5, and a 16.0 $\pm$ 1.0 mm working distance\footnote{The unusually large working distance of 16 mm was used the accommodate the oversized 25.4 mm planchet selected to mitigate the risk of radiological contamination inside the SEM.}.

\subsection{Analysis methods}

The present research seeks to answer two questions by analyzing the generated \puo{} samples:
\begin{enumerate}
    \item What (if any) process parameters cause a statistically significant effect on the morphology of the resulting \puo{} particles?
    \item What phenomenology is associated with those process parameters that do alter the \puo{} morphology?
\end{enumerate}
Unlike traditional effect detection and analysis of DOEs, morphology is inherently a qualitative or high dimensional concept; therefore existing analysis of variance or test-statistic based methods are not applicable.  Instead, computer vision can provide quantitative descriptors of the morphology of particles, and the novel methods generated for this work can answer the above two questions.

We pursued two different analysis pipelines in parallel, sharing components where possible.  To determine the specific process parameters that effect morphology, we generate a neural network which can predict, with the highest possible accuracy, the set of process parameters which generated a given \puo{} particle micrograph.  The success of this prediction, in the absence of nuisance variables\footnote{We mitigate our vulnerability to nuisance variables by utilizing a validation split of data: we assess our performance only on those data on which the network has not been trained.}, proves the effect of the process parameter on morphology.

To investigate phenomenology related to each process parameter, we also perform a clustering analysis.  At a high level, the clustering methodology groups visually similar morphologies into discrete clusters.  This, combined with the proportions of each cluster present for different settings of each process parameter, can illuminate the effect of that parameter morphology.

 \subsubsection{Data preprocessing} 
\begin{figure*}
    \centering
    \includegraphics[width=\linewidth]{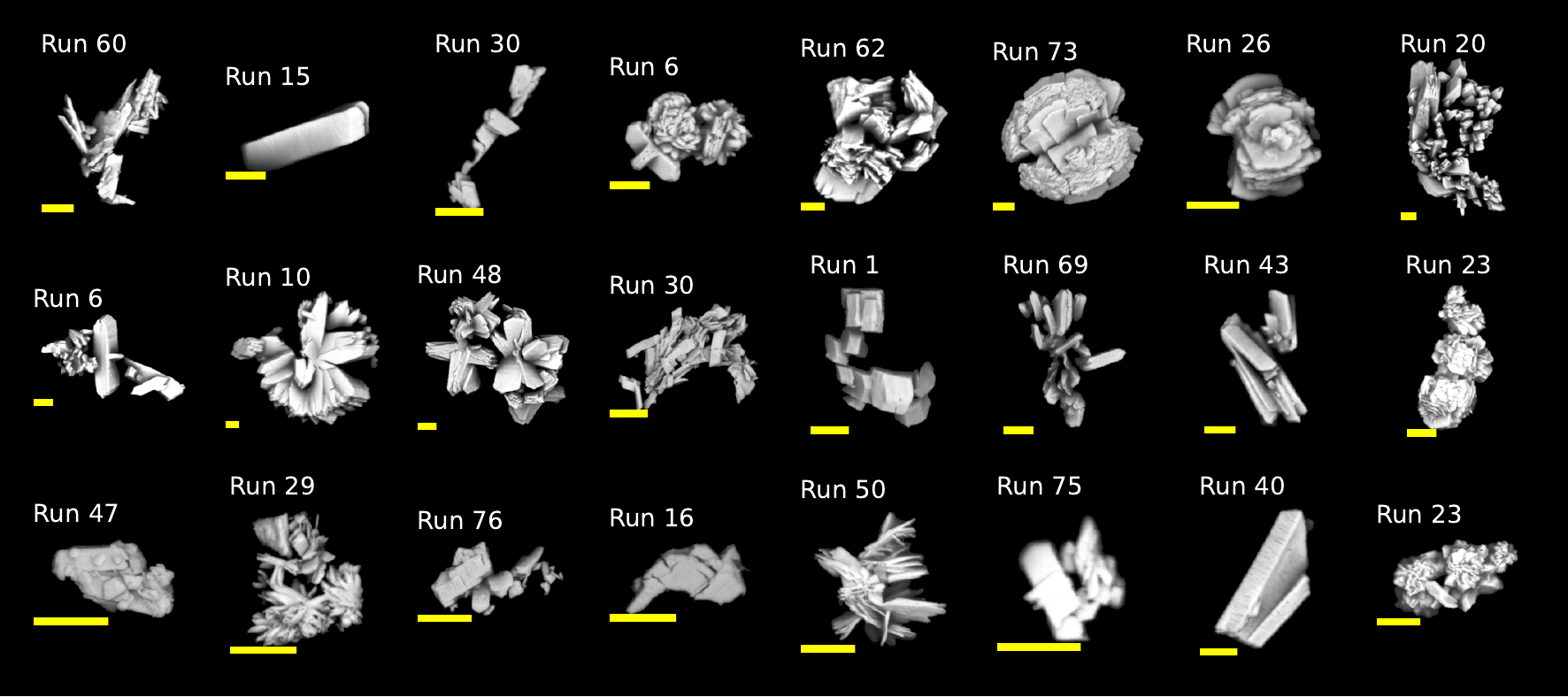}
    \caption{Randomly sampled selection of segmented particles; each scale bar represents 5$\mu$m.}
    \label{fig:particles}
\end{figure*}

Individual \puo{} particles were segmented from SEM images by thresholding; images were cleaned by removing small objects and masking the background regions belonging to the mounting substrate with zeroes. Particles from micrographs acquired at magnifications between 1,000$\times$ and 2,000$\times$ were segmented. Once extracted, any duplicate particles appearing across multiple magnifications and overlapping image regions were identified by size and shape features, then dropped from the dataset. Individual particle segments containing fewer than 10,000 pixels were removed from the dataset before feature extraction. After removing duplicates and making data quality cuts, 10,363 particles remained. A random selection of segmented particles can be seen in Figure~\ref{fig:particles}.

Seven synthetic parameters relating to the plutonium (III) oxalate precipitation process were considered when developing the parameter estimation models. These parameters and their properties were encoded as parameters $A$ through $G$ as shown in Table~\ref{table:enc}; the binary synthetic parameters received encodings from $\{0, 1\}$ and the ternary parameters were encoded from $\{0, 1, 2\}$. The nominal values of two ternary parameters, $D$ and $G$, were used in lieu of the measured values. Whereas seventy-six Pu (III) runs were carried out, there is low confidence in the measured Pu and HNO$_3$ concentration values from the first twenty-five runs, and all particles from those experiments were excluded from the data set.

\begin{table}
\begin{center}
\setcounter{table}{0}
    \caption{Pu (III) oxalate precipitation synthetic parameters and their respective label encodings.}
    \label{table:enc}
    \vspace{1ex}
\begin{tabular}{|c c | c c c|} 
 \hline
 & Parameter & Encoding: 0 & 1 & 2 \\ [0.5ex] 
 \hline\hline
 A & Oxalic acid feed & 0.9M solution & solid & - \\ 
 B & Addition time (min.) & 0 & 20 & 40 \\
 C & Digestion time (min.) & 40 & 20 & 0 \\
 D & HNO$_3$ concentration (M) & 1.0 & 2.0 & 3.0 \\
 E & Strike order & reverse & direct & - \\
 F & Precipitation temperature ($\degree$C) & 30 & 50 & - \\
 G & Pu concentration (g/L) &  10 & 30 & 50 \\ [0.5ex] 
 \hline
\end{tabular}
\end{center}
\end{table}

 \subsubsection{Feature extraction}  Latent representations of the \puo{} particles were extracted using a VQ-VAE model that had previously been trained on uranium oxide SEM images, as described by \cite{girard2021uranium}. After feature extraction by the VQ-VAE, the dataset was split such that 75\% of VQ-VAE embeddings (7,977 particles) were used for the training data with 25\% (2,659 particles) remaining for the validation set.

 \subsubsection{Bottleneck network} 
\begin{figure}
    \centering
    \includegraphics[width=\linewidth]{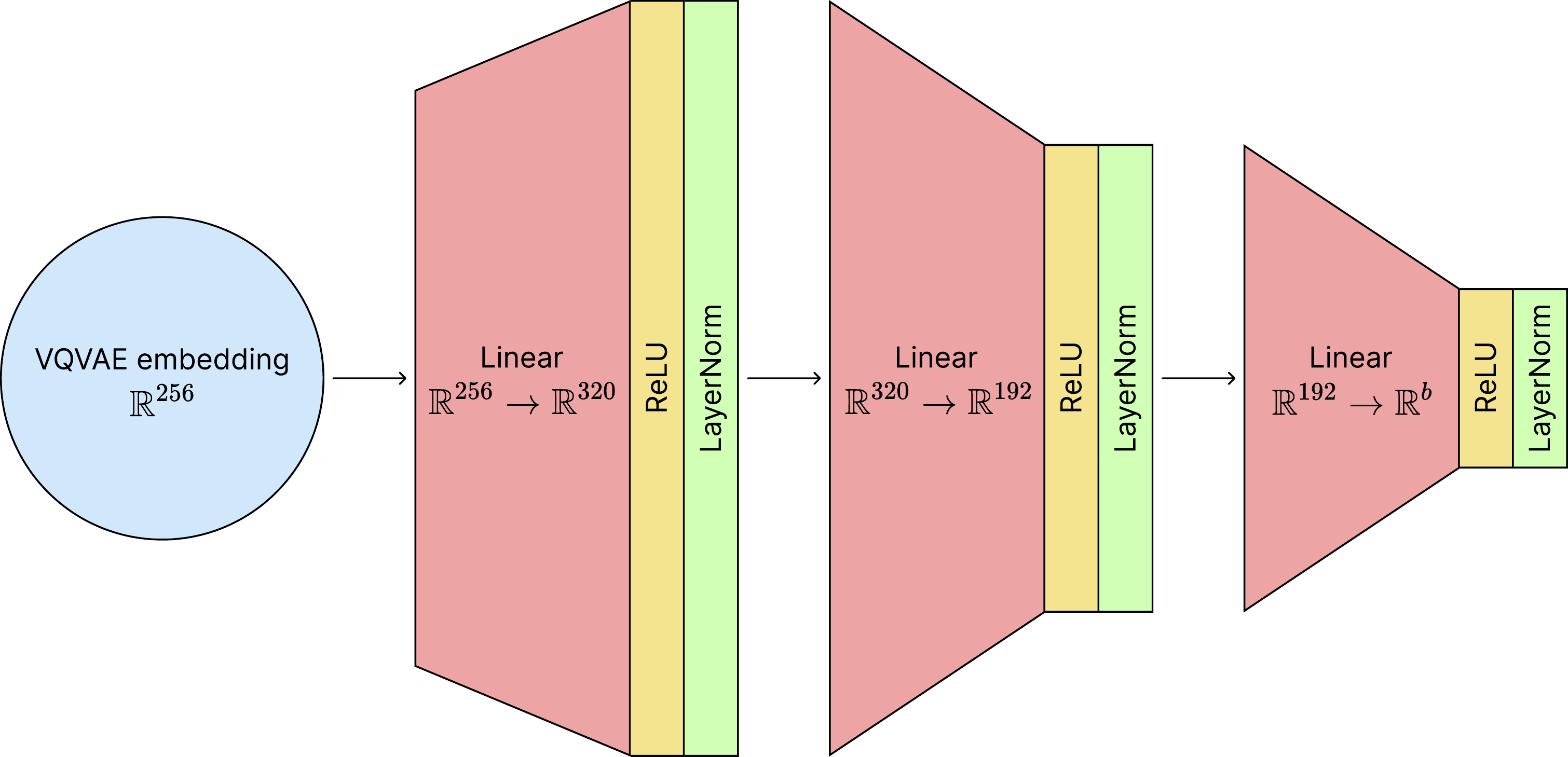}
    \caption{The feed-forward network we use to create a task-specific dimension-reduced encoding of each particle.}
    \label{fig:mlp}
\end{figure}

A simple feed-forward network, depicted in Figure~\ref{fig:mlp}, was used for dimensionality reduction of the VQ-VAE embeddings and process parameter estimation. It consists of three fully connected layers that each apply a linear transformation to the data. We construct the network in the shape of a funnel, such that the dimensionality is reduced a bit at each of the later steps. Empirically, we have found that using hidden layer sizes of 320 and 192 works well, so the layers transform the data according to $\mathbb{R}^{256} \to \mathbb{R}^{320} \to \mathbb{R}^{192} \to \mathbb{R}^{b}$, where $b$ is the ``bottleneck size'' and the dimensionality of the resulting embeddings. After each of the linear layers, we apply ReLU and the LayerNorm-simple algorithm \cite{xu2019understanding}.

The output of this network is a dimension-reduced embedding for each particle. This dimension reduction technique is \textit{task-specific}. Through training on the task, the output embeddings are close to optimal at representing which information from the VQ-VAE embedding is relevant for parameter prediction, and the network is free to discard any additional information.

 \subsubsection{Multi-task training} 
Our overall task is to learn which parameters effect the output morphology of the \puo{} particles.  To learn optimal embeddings for this task, we construct a network which transforms a particle VQ-VAE embedding $\in \mathbb{R}^{256}$ into a transformed embedding $\in \mathbb{R}^{b}$ for some dimensionality $b$. After this network, we use seven linear layers -- each from $b$ dimensions to three dimensions -- in parallel to convert the transformed $b$-dimensional embedding into log-likelihoods for each of the three possible values of the process parameters. For binary process parameters, we manually set the log-likelihood of the third value to $-\infty$.

When training, we use the VQ-VAE embeddings for the particles as input data\footnote{By using the VQ-VAE data as input, we do not propagate gradients to the VQ-VAE itself. This preserves the VQ-VAE model and proves that the model is general use.}. We then apply cross entropy loss for each of the process parameters as our objective function. More specifically, suppose we have the VQ-VAE embeddings for a batch of $N$ particles stacked into a tensor $X \in \mathbb{R}^{N \times 256}$. Our network produces likelihoods for each particle $\mathcal{L}_i (p = v)$ for all process parameters $p$ and their allowed values $v$. Suppose $y_i^p$ is the true value of process parameter $p$ for the $i^{th}$ particle. Then we compute the cross entropy loss for a process parameter $p$ by:

\begin{equation}
CE_p = -\frac{1}{N} \sum_{i = 1}^{N} \log \left( \frac{
    \exp[ \mathcal{L}_{i} (p = y_i^p) ]
}{
    \sum_{v} \exp[ \mathcal{L}_{i}(p = v) ]
} \right)
\end{equation}

We combine the seven resulting cross-entropy values into a single loss by a weighted sum, with a weight of 1.5 for binary parameters and 1.0 for ternary parameters.

\begin{multline}
    CE = 1.5 (CE_A + CE_E + CE_F) + 1.0 (CE_B + CE_C + CE_D + CE_G)
\end{multline}

We then seek to minimize this objective function using the AdamW optimizer. We usually use a learning rate in the range $10^{-4}$ to $10^{-3}$, and we use weight decay with a strength of 0.1 for the network layers proceeding the task layers; weight decay was not used on the task layers. We typically train for several thousand epochs, stopping early if the network begins to overfit to the training data.

 \subsubsection{Classification on clustering} \label{sec:class-clustering}
While the above methodology provides close to optimal parameter estimation and can show the definite effect of process parameters on morphology, it does so in a way that is hard to interpret and may be highly nonlinear. Therefore, we developed a method for performing parameter estimation which will more readily provide phenomenological observations. To do this, we first force the network to ``cluster'' morphology into discrete types, and then use only the cluster membership as information for further prediction. In other words, we estimate process parameters based solely on the morphology \emph{type}.  To do so, we created a Hierarchical Vector Quantization (HVQ) layer to be utilized in our network.

The HVQ layer replaces the continuous, $\mathbb{R}^{256}$ embeddings of a particle with the closest discrete embedding chosen from a two layer hierarchy of traditional vector quantizers.  To achieve this, first the continuous embedding is associated with the closest embedding from the $n_{vq1}$ sized, top-layer ``codebook'' -- the index of this codebook is cached in memory and that index is called the particle \emph{type}.  Then, the continuous embedding is replaced from the closest embedding in the second-layer, $n_{vq2}$ sized codebook associated with the cached top-layer index.  We call the index of the embedding in the second layer codebook the \emph{subtype}.  Finally, the closest second layer embedding is returned as the layer output, and the gradient of the original continuous input is returned as the gradient of the HVQ layer\footnote{While returning the gradient of the input is only an approximation to the (discontinuous) gradient through the HVQ layer, we find that it is sufficient for training our networks.}.

With this setup, the only information available to the task specific layers after the HVQ is that information which has been discretized into particle \emph{types} and \emph{subtypes}. In other words, the prediction layers must assign particle types and subtypes which are descriptive of specific values of process parameters.  Paired with small choices for $n_{vq1}$ and $n_{vq2}$ (we choose $3$ and $4$ for $12$ total possible subtypes), a researcher is enabled to compare morphologies and hypothesize about their cause.

Related works include the standard k-means clustering algorithm \cite{steinley2006k} and several clustering methods built around the VQ-VAE architecture. These methods typically either train a VQ-VAE to use the latent space representations with generic clustering algorithms \cite{chen2022cancer, li2022multi} or use the VQ-VAE's codebook for unsupervised unit discovery \cite{chorowski2019unsupervised, tjandra2019vqvae, van2020vector}. To the authors' knowledge, the use of a vector quantization layer trained on task-specific objectives for a separate clustering task has not yet been described in the literature.


 \subsubsection{Multiple particle parameter estimation} 
Finally, we do not expect that all particles from a given process will be of a single morphology, and so we soften our assumptions to admit the possibility of process parameters affecting the \emph{distribution} of particle morphologies.  Testing this requires only to test the accuracy of parameter prediction on multiple particles at once.  To enable this, we use a very similar neural network architecture as in ~\ref{sec:class-clustering} with an additional sample aggregation layer.

We require that our sample aggregation layer be \emph{permutation invariant}, i.e. that the order of particles presented to the network does not change its output.  In order to do this, we use a Transformer network in conjunction with the classification token (\texttt{cls\_token}) method \cite{vaswani2017attention}, which receives as input a sequence of embeddings from the bottleneck layer of the network, and outputs a single embedding describing the entire sequence.  By not providing any positional information about the sequence, the Transformer is explicitly permutation invariant.  All other aspects of this analysis are identical to the single particle analyses described above.

\section{Results}

\subsection{Parameter estimation}

\begin{figure}
    \centering
    \includegraphics[width=0.5\linewidth]{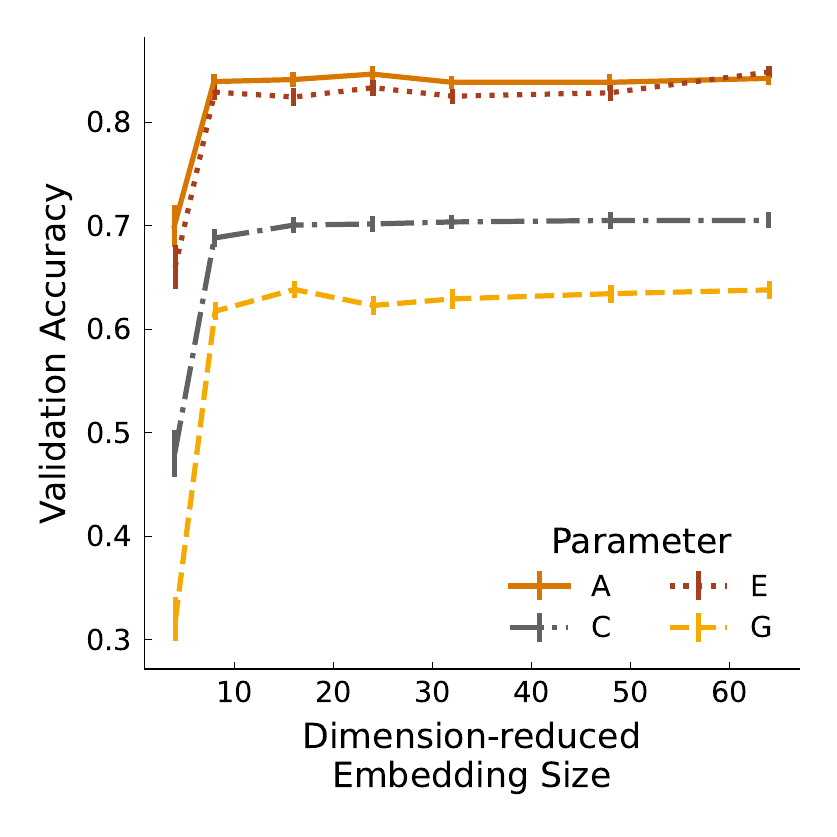}
    \caption{Classification accuracies for selected parameters using single particle samples and varying bottleneck sizes. Increasing bottleneck size leads to an initial increase in accuracy until a plateau is reached.}
    \label{fig:bn_size}
\end{figure}

Classification accuracies for several process parameters using individual particles can be seen in Figure~\ref{fig:bn_size}. For parameter estimation on $C$ (digestion time) and $G$ (Pu concentration) classification accuracies increase sharply with increasing bottleneck size until plateauing around bottleneck sizes of twenty-four; the maximum single particle accuracies for digestion time and Pu concentration were 70.5 $\pm$ 0.8\% and 74.1 $\pm$ 0.8\%, respectively.  Parameters $A$ (oxalic acid feed) and $E$ (strike order) had much higher accuracies overall -- single particle maxima of 84.7 $\pm$ 0.8\% and 84.9 $\pm$ 0.7\%, respectively -- and showed relatively little increase in accuracy when predictions were made by model architectures with larger bottleneck sizes.

The accuracies for the excluded process parameters ($B$, addition time; $D$, HNO$_3$ concentration; and $F$, precipitation temperature) can be seen in Supplementary Figure A.1. Addition time and digestion time, as parameters with inversely correlated encodings, saw similar single particle accuracies across all bottleneck sizes; this trend was seen across all classification experiments with and without aggregated samples. After $A$ and $E$, single particle classification accuracies were next highest for precipitation temperature, another binary parameter. Single particle parameter estimation for HNO$_3$ concentration resulted in classification accuracies similar to the other ternary parameters.

\subsection{Clustering analysis}

\begin{figure*}
    \centering
    \includegraphics[width=\linewidth]{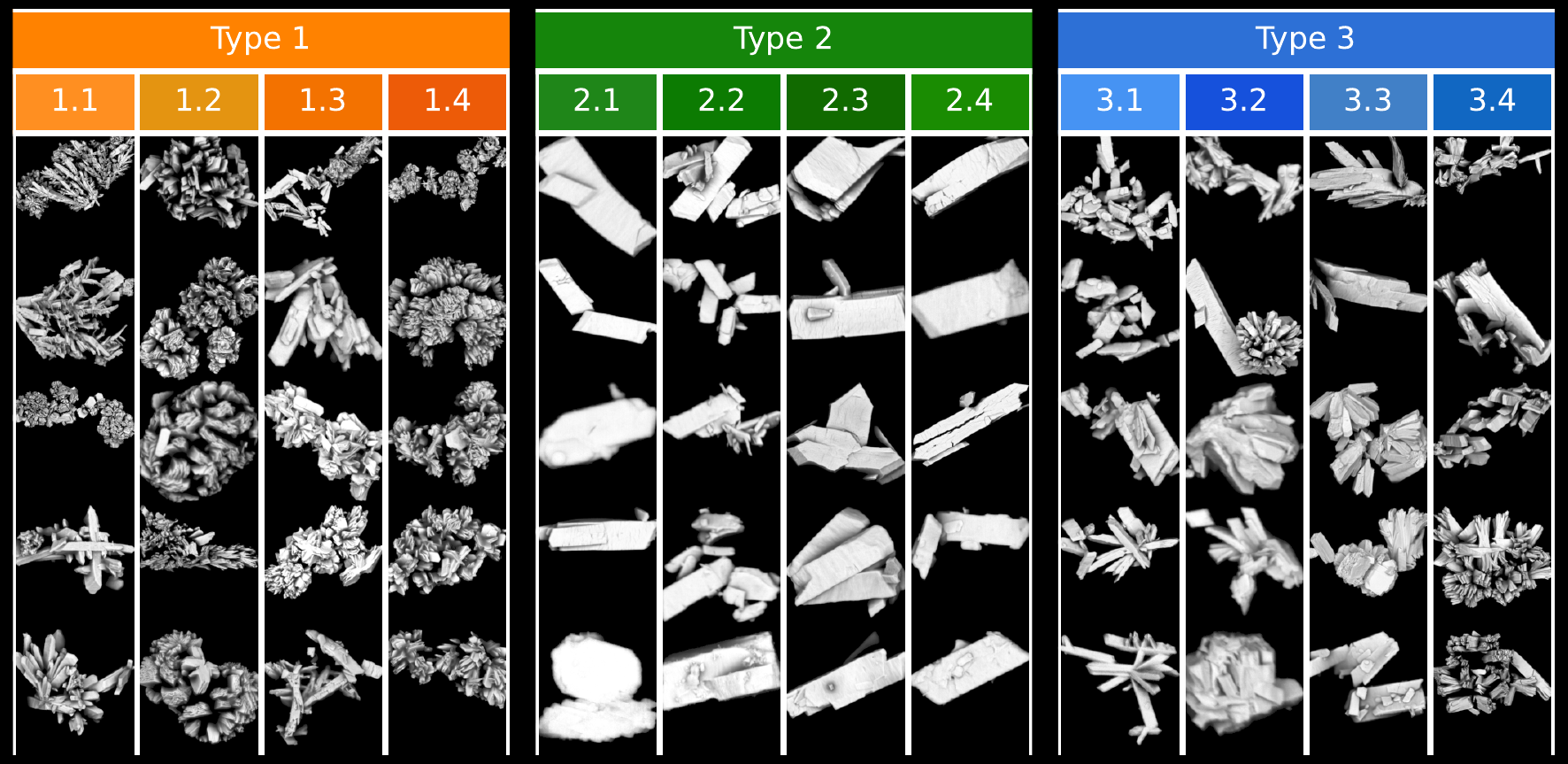}
    \caption{Five example particles nearest to the cluster center for each particle type and subtype learned by the model.}
    \label{fig:cluster_tree}
\end{figure*}

Qualitatively good clustering results were seen across a wide range of choices for HVQ size, though the clusterings of $n_{vq1}=$ 3 and $n_{vq2}=$ 4 were especially useful for particle characterization. A visualization of these results can be seen in Figure~\ref{fig:cluster_tree}. Here, we can see a high degree of visual similarity within each coarse-grained particle type with minimal overlap between the types. The learned particle subtypes share similarities within their parent cluster, but for the most part also have characteristics unique to that subtype. 

The simplest \puo{} particles were placed in the second index (Type 2) by the HVQ. Within the Type 2 group, particles are generally columnar or elongated platelets. Subtype 2.2 shows smaller agglomerations of these morphologies. Subtype 2.3 contains some layered ``wheat sheaf'' particle morphologies. Subtypes 2.1 and 2.4 both show very simple particles.

Type 3 contains more complex particle morphologies than Type 2. The particles clustered into Type 3 are typically either loose agglomerations of simple particle types or particles with ``stellated'' morphologies. Subtype 3.1 shows both of these together, with both loose agglomerations of simple platelets and stellated elongated platelets represented in the five particles nearest the codebook center. Stellated columnar particles are seen for subtypes 3.2 and 3.3. Subtype 3.4 features columnar particles that show ``delamination'' or ``fraying'' at the particle ends.

The most complex particles and agglomerates were assigned to Type 1 by the HVQ model. The particles seen here are qualitatively much more complex than those assigned to Type 2. Irregular ``dendritic'' particles and compact ``rosettes'' are the most prominent morphologies present.

\begin{figure}
    \centering
    \includegraphics[width=0.5\linewidth]{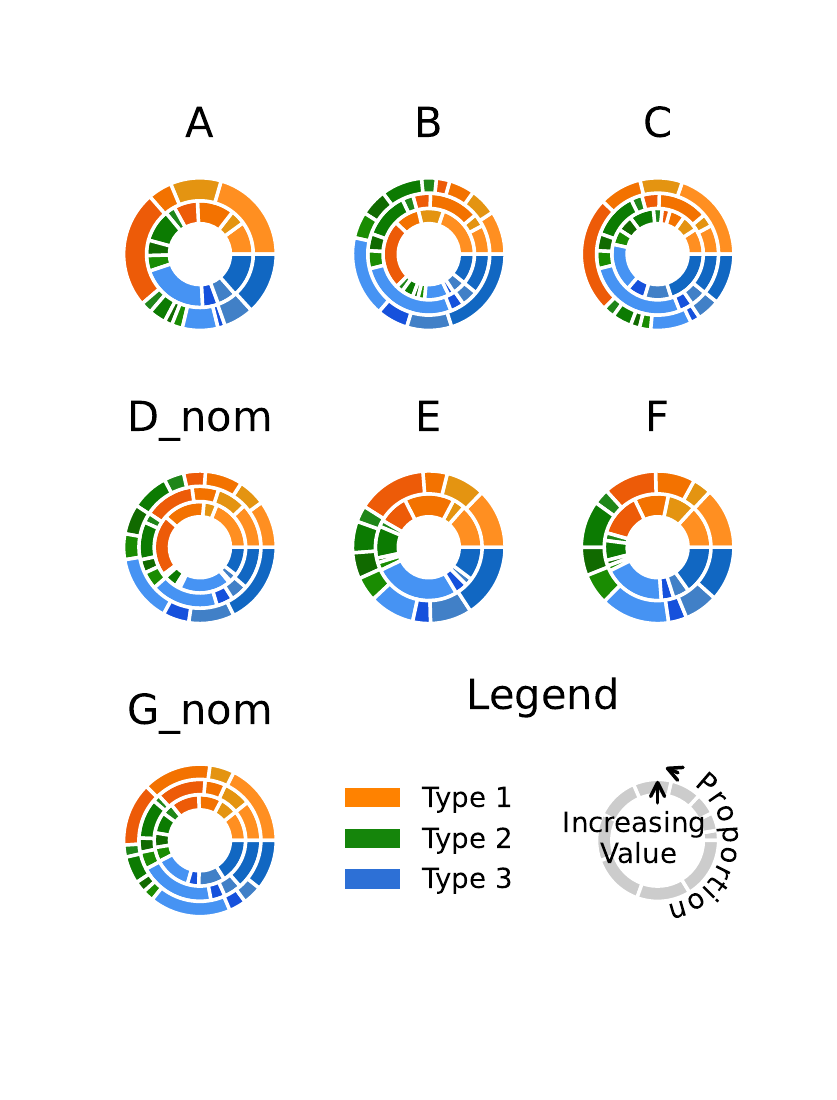}
    \caption{Proportion of particle types and subtypes represented by color around the circumference versus value of parameter with increasing radius.}
    \label{fig:cluster_proportions}
\end{figure}

To discover which parameter settings create which particle types, Figure~\ref{fig:cluster_proportions} plots the distribution of particle types and subtypes for each parameter and setting. The inner rings for each parameter represent the lower setting in Table~\ref{table:enc} with each subsequent outer ring representing the next setting. For example, the inner ring of parameter $B$ is a 0 minute addition time, the middle ring is a 20 minute addition time, and the outer ring is a 40 minute addition time.

Parameter $A$, oxalic acid feed, shows one of the largest effects on the resulting particle morphologies. The low setting, representing a 0.9 molar oxalic acid feed solution, produces primarily Type 3 moderately complex particles. When using a solid oxalic feed the proportion of very complex Type 1 particles nearly doubles from 32\% to over 61\%; the proportion of simple Type 2 particles halves from 22\% to 10\%.

For addition time ($B$) and digestion time ($C$) more complex particles are seen with immediate feed additions and longer digestion times. Approximately 90\% of particles produced under these conditions were assigned either Type 1 (63\%) or Type 3 (26\%). For 40 minute feed addition times and and no digestion of the precipitates, the proportion of moderately complex Type 3 particles increases to about 50\%, with a huge decrease in very complex particles seen and a slight increase for simple particle morphologies, each representing about 25\%.

The nominal nitric acid concentration ($D$) saw more Type 1 particles at lower acid concentrations and more simple and moderately complex particles particles at higher acid concentrations. The nominal plutonium concentration ($G$) saw the opposite trend: more complex particles at higher Pu concentrations and more simple particles at lower concentrations. More particles produced at higher precipitation temperatures ($F$) were clustered as simple Type 2 particles. For strike order ($E$), the proportion of particles remained consistent for either strike order with approximately 40\% Type 1, 40\% Type 3, and 20\% Type 1.

\begin{figure*}
    \centering
    \includegraphics[width=\linewidth]{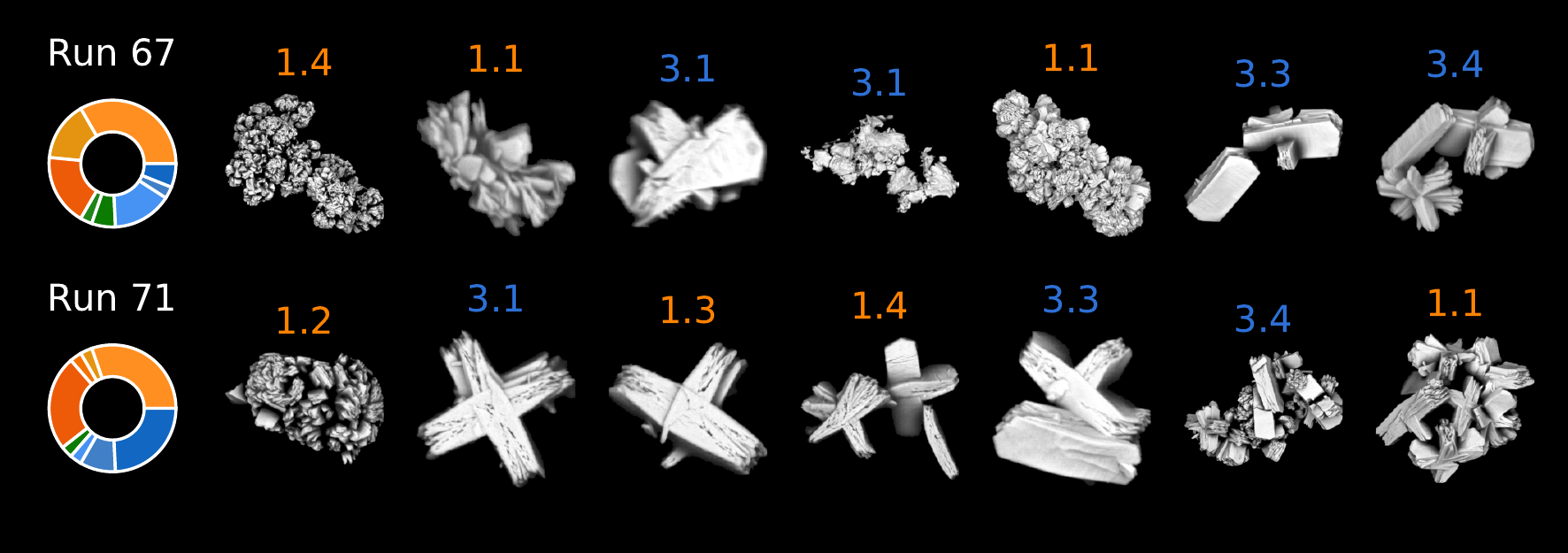}
    \caption{Distribution of clustered particle types and randomly selected example particles for duplicate experimental runs 67 and 71.}
    \label{fig:dupesA}
\end{figure*}

Particle type distributions were also calculated for each experimental run (Supplementary Figures A.2 and A.3). From this we can begin measuring the distribution of particles produced by each set of experimental parameters. In the original seventy-six runs there were several sets of duplicate experiments that allowed for the consistency of the methods to be evaluated. In most -- but not all -- cases the duplicated experiments had similar distribution of particles assigned by the HVQ. One such instance of good accordance between duplicate is shown in Figure~\ref{fig:dupesA}. Runs 67 and 71 each had about two-thirds of particles assigned to Type 1; these particles show a complex rosette morphology. The remainder of particles for runs 67 and 71 were primarily categorized as Type 3 and typically showed columnar ``cruciform'' morphologies, though these morphologies were sometimes categorized as Type 1, particularly when present in agglomerations.

\begin{figure*}
    \centering
    \includegraphics[width=\linewidth]{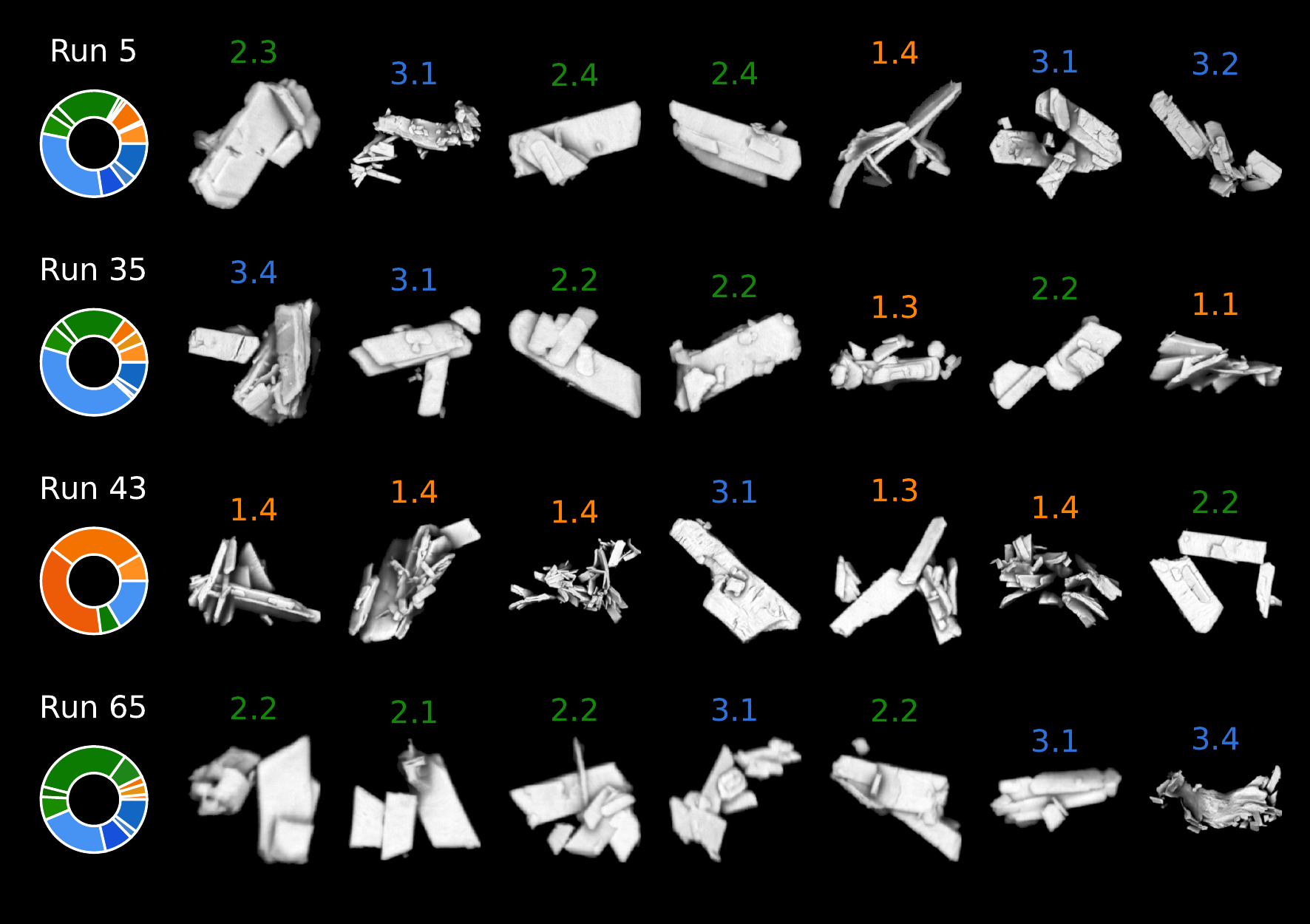}
    \caption{Distribution of clustered particle types and randomly selected example particles for duplicate experimental runs 5, 35, 43, and 65.}
    \label{fig:dupesB}
\end{figure*}

Figure~\ref{fig:dupesB} shows the distributions and example particles for duplicate experimental runs 5, 35, 43, and 65. Runs 5, 35, and 65 show similar particle distributions, with Types 2 and 3 combining to make up the majority of particles and few Type 1 particles. Run 43, on the the other hand, has a majority of its particles categorized as Type 1. The crystal habit of particles in these duplicate runs are consistently elongated platelets. Runs 5, 35, and 65 generally formed smaller loose agglomerates of simple particles or ``wheat sheaf'' morphologies. Larger loose agglomerates are seen in much higher frequencies for the run 43 particles, which is a likely cause for the shift in particle distributions towards the more complex Type 1. 

\subsection{Multi-particle parameter prediction}

\begin{figure}
    \centering
    \includegraphics[width=0.5\linewidth]{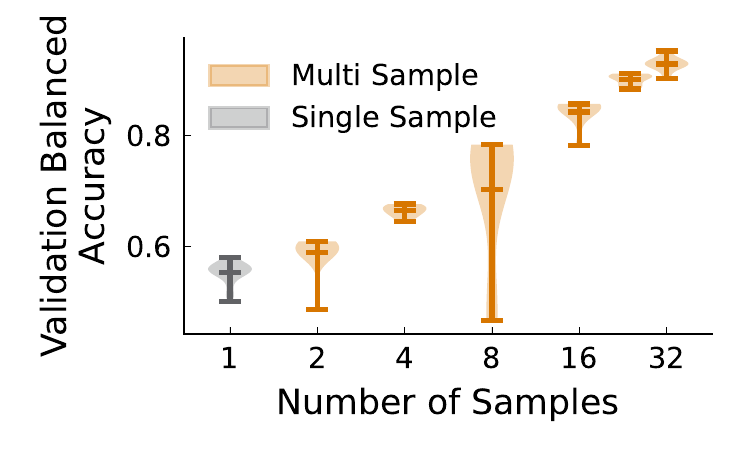}
    \caption{Average classification accuracy with increasing sample size for multi-particle parameter prediction. Uncertainty is shown as the distribution of resultant accuracy after each of 10 full retraining with varying random states.}
    \label{fig:n-samples}
\end{figure}

Overall, multi-particle parameter prediction yielded superior performance to that of single particle parameter prediction.  As shown in ~\ref{fig:n-samples}, the average classification accuracy across all parameters increases with increasing sample size, achieving $>$ 90\% prediction at 32 particles in each sample.

Sampling only within certain particle size quantiles led to a marked degradation in performance compared to uniformly sampling particles from an experimental run across all quantiles together. For some some parameters, however, interesting relationships between synthetic process parameters and particle properties begin to emerge from the quantile samples. The relationships we explored were specifically in terms of the quartiles, Q1-Q4, of the particle size.

\begin{figure}
    \centering
    \includegraphics[width=0.5\linewidth]{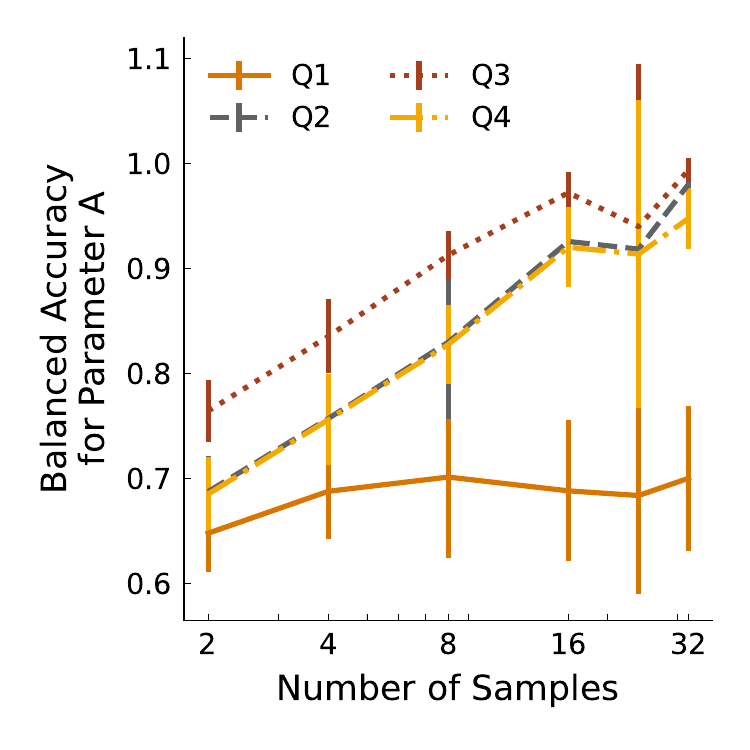}
    \caption{Sampling particles exclusively from a single size quantile led to a decrease in classification accuracy when compared to sampling from all quantiles. Parameter estimation on oxalic acid feed saw much higher accuracies when using particles from any quantile but the first quantile, Q1.}
    \label{fig: quantiles}
\end{figure}

Parameter estimation for oxalic acid feed saw diverging results with Q1 and Q2-Q4 initially having similar classification accuracies at small sample sizes (Figure~\ref{fig: quantiles}). With two particles available for classification, the accuracies were 65.0 $\pm$ 3.2\% and 68.3 $\pm$ 3.9\% for the first and fourth quartiles, respectively -- significantly lower than the 81.2 $\pm$ 0.7\% seen when sampling uniformly. As more particles were aggregated the classification accuracy increased when  particles from Q2-Q4 were sampled, outperforming single particle accuracies with $N>$ 8. No improvement was observed with increasing sample size when the small Q1 particles were sampled.

The results for all parameter estimation with quantile-split particles can be seen in the Supplementary Figure A.4. No significant divergence between Q1 and Q4 can be seen for nitric acid concentration, strike order, precipitation temperature, and Pu concentration, and all quantiles remain below the accuracies when sampling from all quantiles. Predictions on addition and digestion time show slightly better performance using Q1-Q3 particles over Q4 particles with more particles aggregated. Predictions on Pu concentration shows the opposite trend of that for oxalic acid feed, with prediction on Q4 particles showing no significant improvement with increasing sample size.

\section{Discussion}

The implementation of a bottleneck network for dimensionality reduction of \puo{} SEM data proved to be capable and versatile. Bottleneck sizes in the range of $b=$ 8 to $b=$ 16 provided adequate performance on both parameter estimation and clustering tasks, though slightly larger bottleneck sizes than this were preferred for parameter estimation. Training the bottleneck model was stable over a wide range of bottleneck sizes with and without the HVQ clustering layers.

Estimation of the seven synthetic parameter, ranged from 55\% to 82\% accurate when a single particle was used for classification. Sampling particles from only within a single size quantile led to a general decrease in performance when compared to uniformly sampling particle. However, certain parameters showed opposite trends on single quantile, multiple particle predictions. Larger particles led to higher accuracies for parameter $A$ than smaller Q1 particles, but the same Q4 particles led to much lower accuracies for several parameters, including $E$ and $G$. While this finding reveals that certain particle morphologies and conglomerates may be more useful for certain parameters, it also reveals the need for representative sampling methods that eliminate biases. On the data science side, the very smallest particle segments, also known as ``fines,'' were removed from the data set before training models for data quality reasons -- segmented fines tended to either be blurrier or have jagged edges. On the experimental side, it was observed that some \puo{} particles would fall out of the isopropyl alcohol slurry before a sample could be aliquoted for SEM imaging. It is assumed that the largest particles and conglomerates would be the first to settle out of the slurry. Both of these present sources of biases that would affect a trained model's behavior and, possibly, its ability to generalize to new data from different sources.

Information about Pu (III) oxalate phenomenology could be learned from the classification and HVQ clustering results. The clustering analysis was useful for finding and categorizing particles and relationships between synthesis parameters and the resulting surface morphologies could be discovered. Firstly, the oxalic acid feed, which was either a 0.9 M solution or a solid, saw high classification accuracies and had a large effect on the resulting particle distributions. Solid oxalic acid feeds produced significantly more very complex Type 1 particles than aqueous feeds, which favored moderately complex and simple morphologies.  Additionally, addition and digestion time (which are complementary in the designed experiment analyzed) show clear evidence of being a ``threshold" variable: the particle type distribution of the lowest value for addition time is different than the relatively very similar particle type distribution at the higher two values.

The strike order process parameter was able to achieve classification accuracies that were $>$5 percentage points higher than other process parameters, indicating that the selection of this binary parameter -- direct or reverse -- produces a very machine learn-able change in the resulting surface morphology characteristics. However, the clustering analysis using the HVQ did not show a shift in particle type distribution for strike order despite the expectation that direct strikes would produce more complex particle types. Smith et al. noted ``more agglomeration and aggregation (rosettes, spherulites, and laminar aggregates)'' from direct aqueous strikes \cite{smith1976effect}, compared to the reverse strike which formed ``monolithic lath particles and loosely bound agglomerates'' \cite{burney1984controlled}. Many parameters, however, are difficult to disentangle from the others. For solid feeds, the precipitations were always carried out with a fixed strike order (direct), addition time (0 minutes), and digestion time (40 minutes). In contrast, aqueous oxalic acid feeds could be added to the Pu nitrate or vice versa with any addition and digestion time. Only considering direct strikes and immediate addition times, particles from solid oxalic acid feeds tended to produce more rosettes and stellated complex particles than aqueous feeds, which produces fan-like and dendritic complex particles.

Comparing duplicate experimental runs revealed that the morphological characteristics of particles produced under a certain set of conditions were generally repeatable, with the HVQ clustering analysis showing agreement between the clustering distributions of these runs. It also revealed aspects of the HVQ clustering that have room for improvement. First, while the top level types generally had the same distributions, the more fine-grained subtypes did not; particles that were qualitatively very similar would often be assigned different subtypes, as was the case with subtypes 1.1 and 1.4 in Figure~\ref{fig:dupesA}. Second, the HVQ's learned clustering is largely based on the overall morphology of particles and agglomerates/conglomerates rather than the crystal habit of the particles. While the preference of overall appearance over crystal habit was expected, the ability to separately analyze the individual components of agglomerated and conglomerated particles and compare to other individual particles or aggregated components might lead to better clustering results.

To reduce biases in sample preparation, a deeper dive should be taken into how samples are handled, split, and aliquoted for analysis. As changes are proposed and made on the experimental side, it would be prudent to keep data scientists knowledgeable of experimental work and for data analytical methods to be subsequently reevaluated. For example, if changes are made to obtain a more representative sample of particles that includes more large conglomerates the distribution represented in the new data set would have shifted from what was used to train the parameter estimation models. And while parameter estimation was best accomplished by aggregating particles sampled uniformly across the data, a new distribution with more large particles could lead to a decrease in performance for parameter $A$, which did not perform as well with smaller particles.

\section{Conclusions}

The HVQ method was shown to be a robust categorization tool for particle morphology analysis, demonstrating the capability to successfully cluster ex-Pu (III) oxalate \puo{} particles with similar morphological characteristics. Quantification of particle type distributions was rapid and reliable, automating what is a relatively subjective and slow task even when performed by experts. Phenomenological trends between the synthetic parameters and resulting surface morphologies were identified, pointing towards the main effects of the statistical design of experiment study.

Future work should investigate the chemical, physical, and morphological properties of materials synthesized by the Pu (IV) oxalate precipitation route; the methods presented in this work should be expanded to similarly characterize and describe the relationship between Pu (IV) oxalate process parameters. The ability to distinguish between ex-Pu (III) and ex-Pu (IV) oxalate \puo{} particles should be evaluated. Collaboration between subject matter experts with knowledge of physical and chemical processes and data scientists with the computer vision tools to describe and quantify the effects of these processes will continue to be important for future efforts.


\section*{Acknowledgements}

This work was funded by the National Technical Nuclear Forensics Center (NTNFC) within Countering Weapons of Mass Destruction (CWMD) , formerly the Domestic Nuclear Detection Office (DNDO), of the Department of Homeland Security and conducted at Pacific Northwest National Laboratory. Pacific Northwest National Laboratory is operated by Battelle Memorial Institute for the United States Department of Energy under contract DE-AC05-76RL0-1830.

\bibliographystyle{elsarticle-num}
\bibliography{refs}{}

\clearpage
\begin{appendices}
\section{Supplementary figures}

\begin{figure}[h!]
    \centering
    \includegraphics[width=0.5\linewidth]{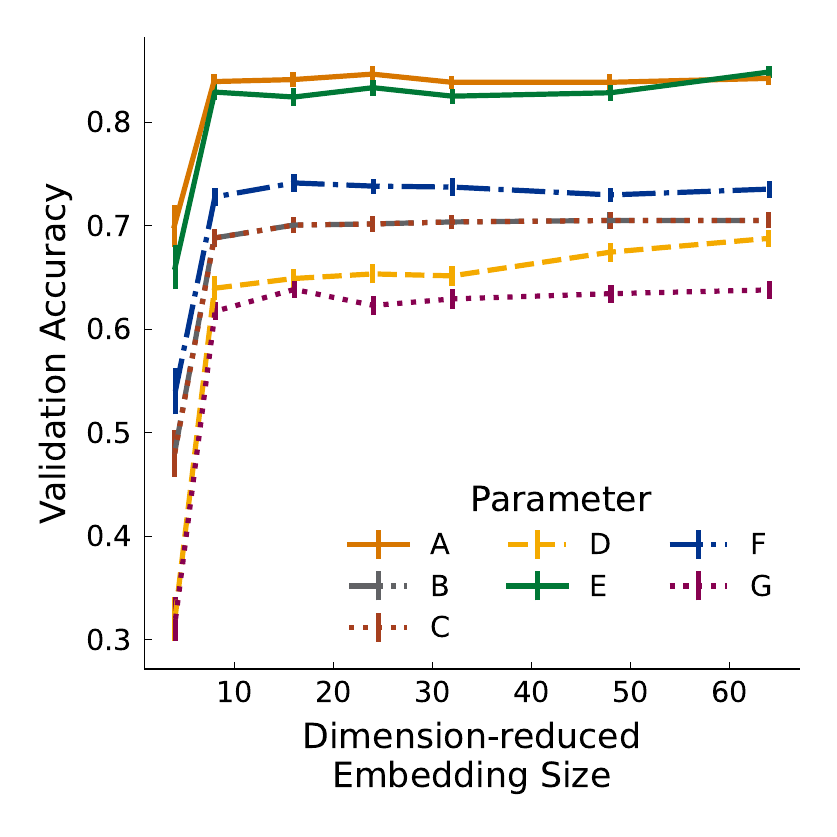}
    \setcounter{figure}{0}
    \renewcommand{\thefigure}{A.\arabic{figure}}
    \caption{Single particle classification accuracies for all process parameters made by classifiers with varying bottleneck sizes.}
    \label{fig:bn_all}
\end{figure}

\begin{figure}[]
    \centering
    \renewcommand{\thefigure}{A.\arabic{figure}}
    \includegraphics[width=\textwidth]{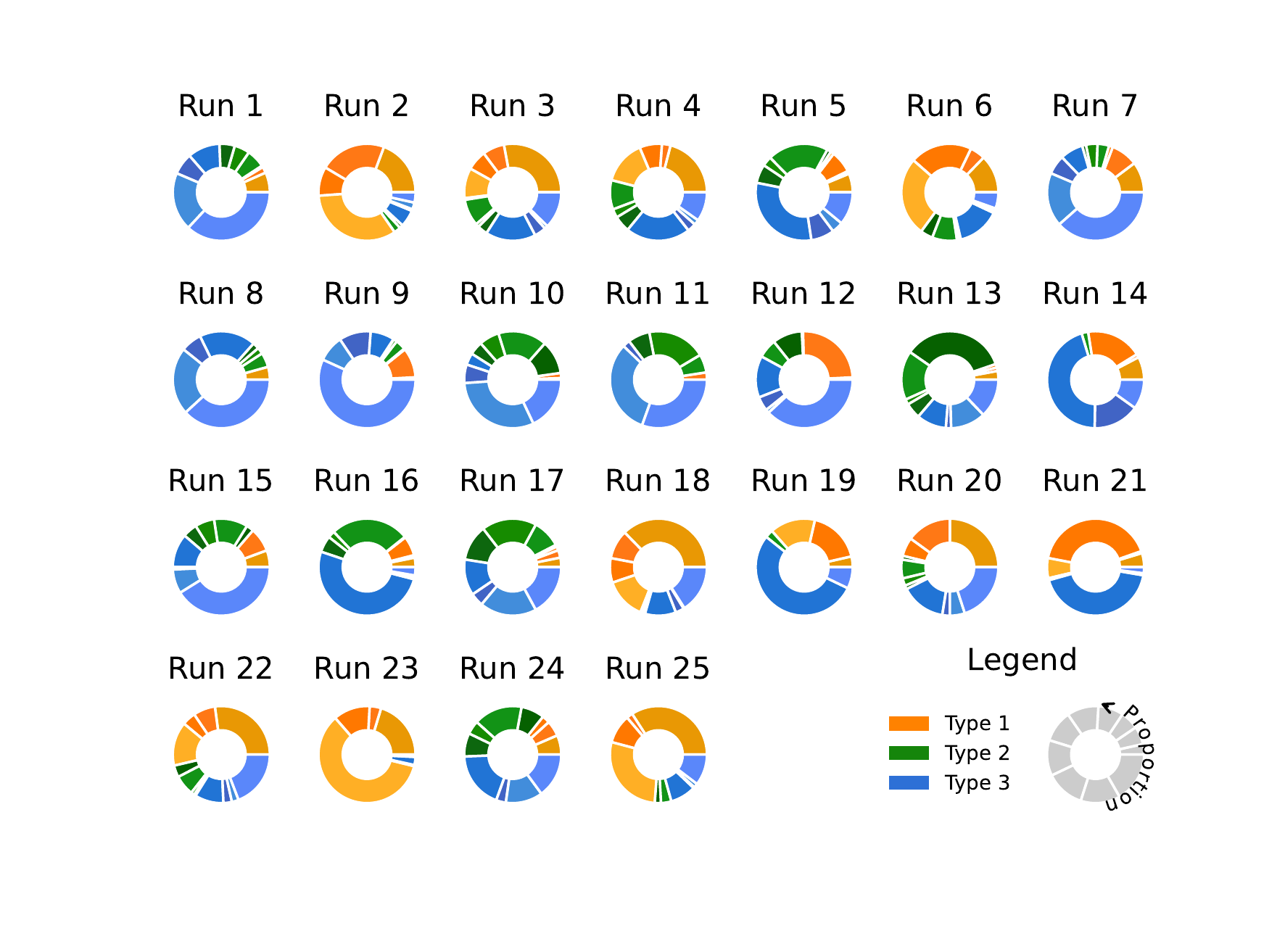}
    \caption{Clustered particle types and subtypes for runs 1-25, which were initially excluded due to low confidence in the HNO$_3$ and Pu concentration measurements.}
    \label{fig:run1-25_props}
\end{figure}

\begin{figure}[]
    \centering
    \renewcommand{\thefigure}{A.\arabic{figure}}
    \includegraphics[width=\textwidth]{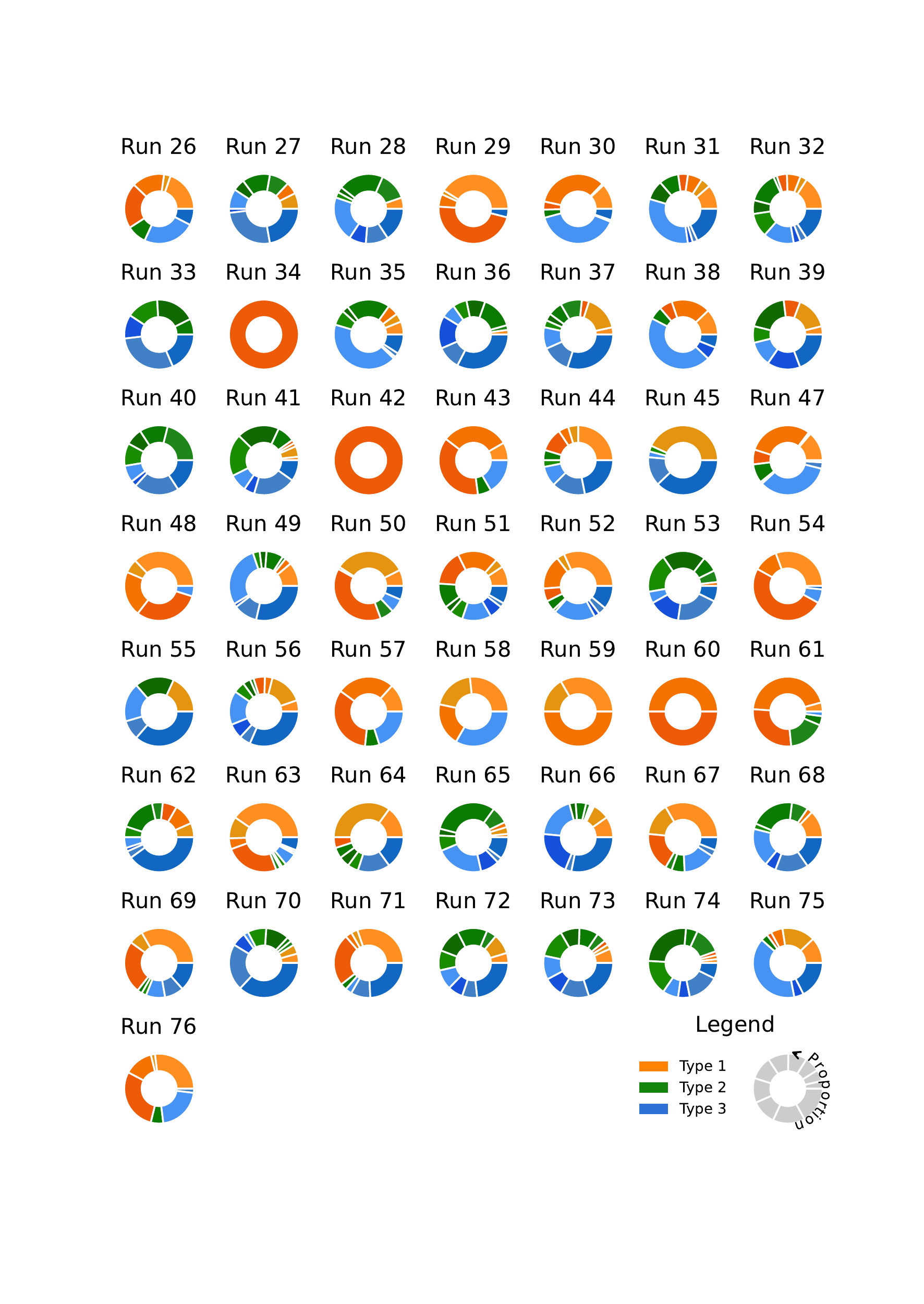}
    \caption{Clustered particle types and subtypes for validation set data from runs 26-76.}
    \label{fig:validation_props}
\end{figure}

\begin{figure}[]
    \centering
    \renewcommand{\thefigure}{A.\arabic{figure}}
    \includegraphics[width=\linewidth]{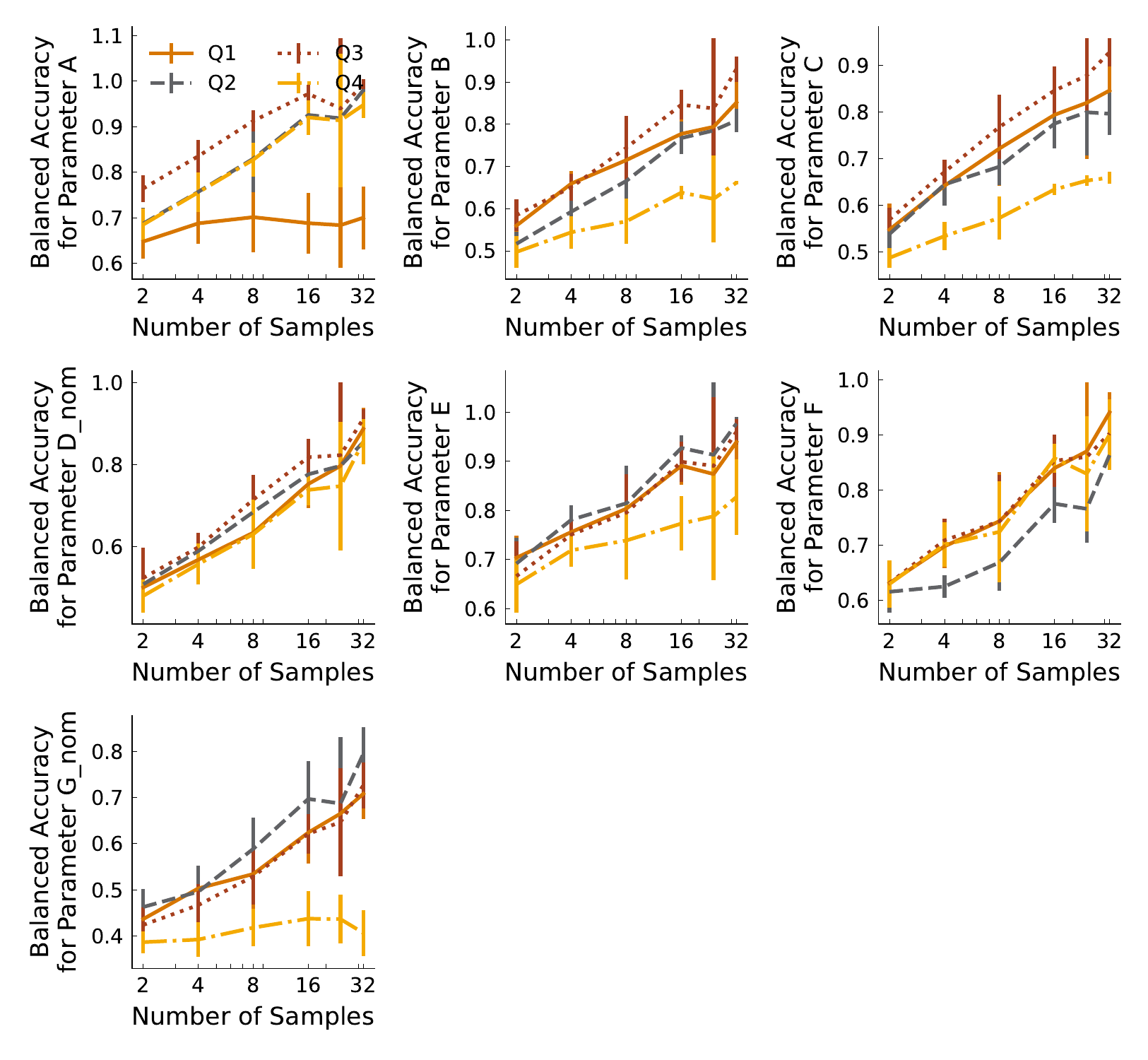}
    \caption{Aggregated particle accuracies from the $b=$ 24 model sampling from all particles uniformly and within each size quantile.}
    \label{fig:bn24_quantiles}
\end{figure}
\end{appendices}

\end{document}